\documentclass{ws-procs975x65}

\begin{document}

\title{Microlensing events from galactic globular clusters}

\author{Philippe Jetzer}

\address{
Institute of Theoretical Physics\\
           University of Z\"urich, 
Winterthurerstarsse 190,CH-8057 Z\"urich, Switzerland\\
E-mails: jetzer@physik.uzh.ch}


\abstract{
We present an analysis of the large set of microlensing events detected
so far toward the Galactic center with the purpose of investigating whether some 
of the dark lenses are located in Galactic globular clusters.
We find that in four cases some events might indeed be due to lenses
located in the globular clusters themselves.
We also give a rough estimate for the average lens mass
of the events being highly aligned with Galactic globular cluster centers
and find that, under reasonable assumptions, 
the deflectors could most probably be either brown dwarfs, M-stars or stellar remnants.}

\section{Introduction}

Globular cluster could contain a sizeable amount of dark matter in form of
brown dwarfs or low mass stars. This is still an open issue and a possible way to test 
this is to use microlensing observations Refs.~\citen{pacz94,luca,straessle}. The idea is to monitor
globular clusters in front of rich background fields of stars
of the galactic bulge. 
In this case, when the lens belongs to the cluster population,
its distance and velocity are roughly known. This way it is possible to get 
a more accurate estimate for the lens mass. 
Such a study has already been performed ~\cite{luca,straessle} and some events were found
which might be associated with lenses in globular clusters Refs.~\citen{straessle,sahu,luca}. 
 
We analysed the possible MACHO content in  
a large set of Galactic Globular Clusters (hereafter GGCs) some of which are highly aligned 
with a non negligible number of microlensing events 
detected toward the Galactic Center (hereafter GC).
The data set included 4697 microlensing events
detected in the last years by the MACHO, OGLE,
and MOA collaborations in direction of the GC.
In our analysis we focused on the configuration in which the lens is hosted in a GGC
and the source is located either in the
Galactic disc or bulge.

\section{Results and conclusions} \label{sec:results}

Aiming at discriminating among events due to lenses hosted either in GGCs
or in the Galactic bulge/disc, 
we first made a rough selection of events being aligned with a GGC.
In particular, for every given GGC, we considered a sphere of radius $r_t$
(corresponding to its tidal radius),
centered at the GGC center,
and we selected, as a first step, only the events being included in one such contour.
By doing so, out of the original 4697 events, we were left with 118.

The GGCs aligned with at least one event are given in Table \ref{GGC1}.
Some properties are also reported, such as the tidal and the core radius, the distance 
from the Sun ~\cite{harris} and the number of included events.
The different core radii and central densities explain
the large variations of the optical depth among the GCCs.

\begin{table}[h]
\tbl{GGCs being aligned with at least one detected microlensing event.
For each of them the number of aligned events $N_{tot}$ and
the corresponding average duration $<t_E>$ (in days) is given.
For every GGC, $r_t$ is the tidal radius (in pc), 
$r_c$ is the core radius (in pc), 
$r_{sun}$ is the distance of its center from the Sun (in kpc)
and $\tau$ is the optical depth toward its center
in units of $f\times 10^{-5}$, $f$ being the fraction 
of dark matter mass in the GGC.
\label{GGC1}} 
{\begin{tabular}{c|cccccc}
\hline             
{Cluster ID} &$r_{sun}$&$r_t$& $r_c$&  $\tau$ & $N_{tot}$& $<t_E>$  \\
\hline                                                 
 Pal 6        & 5.9  &     14.3  & 1.13 & 0.24   &   2    & 76.3    \\
Terzan 9      & 6.5  &     15.5  & 0.06 & 0.78   &   2    & 17.8   \\
NGC 6522      & 7.8  &     37.3  & 0.11 & 6.06   &   36   & 16.8   \\
NGC 6528      & 7.9  &     38.1  & 0.21 & 1.01   &   38   & 25.2   \\
NGC 6540      & 3.7  &     10.2  & 0.03 & 23.74  &   29   & 22.8   \\
NGC 6553      & 6.0  &     14.2  & 0.96 & 1.01   &   7    & 43.0   \\
NGC 6558      & 7.4  &     22.5  & 0.06 & 7.20   &   1    & 24.5   \\
NGC 6624      & 7.9  &     47.2  & 0.14 & 4.43   &   1    & 223.0   \\
NGC 6656      & 3.2  &     27.0  & 1.32 & 0.75   &   2    & 112.7   \\
\hline
\end{tabular}}
\end{table}

Due to the GGC structure, we expect the predicted number of events to
be the largest toward
their centers and to decrease as we move toward their borders.
Since the alignment between an event and a projected cluster contour 
does not assure that the deflector belongs to
the GGC, this alignment possibly being accidental,
we made a further, rough selection
and considered only the events being included in the projected 
contour of a sphere centered at a GGC center and of radius $r_i=2\times r_t/5$
(this including on average $90 \%$ of the total cluster mass).
We then distinguished between $inner$ and $outer$ events, the former being inside $r_i$
and the latter being included in the circular ring of internal radius $r_i$ and outer radius $r_t$.
By doing so, we assumed all the outer events to be due to Galactic 
bulge/disc deflectors (this possibly underestimating the events due to GGC lenses),
whereas we left open the possibility that among the inner events
some could still be attributed to bulge/disc deflectors.
At the end we are left with 28 inner events,
among which 7 (17/4) have been detected by the MACHO (OGLE/MOA) collaboration.

\begin{table}[h]
\tbl{GGCs with inner events.
For each of them $N_{in}$ is the number of events inside a projected radius 
$r=2\times r_t/5$
and, for this subset of aligned events,
$<t_E>$ is the mean Einstein time (in days)
and $<m>$ is the average predicted lens mass in units of
solar masses.
$N_{GGC}$ ($N_{BD}$) is the number of events, out of $N_{in}$, that we expect to be due to GGC (Galactic bulge/disc) lenses.
$\Gamma_{exp}$ is the expected event rate in units of $f\times \mu_o^{-1/2}\times 10^{-3}/year$, while 
$n_{GGC}$ is $N_{GGC}$ per unit area (in $\mathrm{degree^{-2}}$).
\label{GGC2}}
{\begin{tabular}{c|ccccccc}
\hline             
{Cluster ID} &$N_{in}$& $<t_E>$  & $<m>$  &$N_{BD}$    & $N_{GGC}$ &$\Gamma_{exp}$ &$n_{GGC}$\\
\hline                                                          
 NGC 6522     &   8   &   13.1   &  1.63  & 4.1$\pm$ 2.0& 3.9      &  0.66    & 51.4     \\
 NGC 6528     &   7   &   13.0   &  2.98  & 4.9$\pm$ 2.2& 2.1      &  0.09    & 27.5     \\
 NGC 6540     &   7   &   17.2   &  0.06  & 4.2$\pm$ 2.0& 2.8      &  1.56   & 112.3    \\
 NGC 6553     &   4   &   35.7   &  0.62  & 0.6$\pm$ 0.8& 3.4      &  0.08    & 185.4    \\

\hline

\end{tabular}}
\end{table}

An estimation of the predicted 
number of events, $N_{GGC}$, due to MACHOs in a given GGC,
can be roughly made as follows.
Assuming that all the outer events are due to Galactic bulge/disc lenses,
we calculate how many such events, $N_{BD}$, are expected in the inner region of a 
GGC contour assuming that the number of events 
is proportional to the covered area
and that the background source distribution is uniform inside every GGC contour.
Thus we assume that the microlensing rate for Galactic bulge/disc events
is constant over the entire small
area within the tidal radius of the considered globular cluster.
By doing so, $N_{BD}$ is simply proportional
to the monitored area. Clearly, also with these assumptions, which are reasonable,
given the very small area considered, one expects fluctuations in the number
of events in a given area. We assume the
fluctuations to follow Poisson statistics, in which case they are given by
$\sim \sqrt{N_{BD}}$.
By doing so, for every GGC considered,
$N_{GGC}$ turns out to be around 2-4 per cluster (see Table \ref{GGC2}) and in two cases
this number is larger than the estimated fluctuation of $N_{BD}$.
Given these numbers we cannot claim for any clear evidence of
lenses hosted in GGCs. Nonetheless, it is remarkable that for 
the 4 cases considered the value of $N_{GGC}$ is positive
and most probably underestimated, since the assumption that all the events 
lying in the outer ring are due to bulge/disc deflectors possibly overestimates $N_{BD}$.

Assuming that the deflector is a GGC MACHO, we can estimate its mass
through the relation $R_E/t_E=v_r$, where $v_r$ is the 
lens-source relative velocity orthogonal to the l.o.s.,
$t_E$ is the event Einstein time and $R_E$ 
is the Einstein radius. For $v_r$ we adopt the value of the proper motion of the considered
globular cluster as given in the literature.
As reported in Ref.~\citen{harris}, the mean GGC tidal radius is of the order
of tens of pc,  this making the GGC extension relatively small compared to 
the average lens distance from Earth or the source distance 
(of the order of kpc),
since we are assuming Galactic bulge/disc sources and the GGCs are kpcs 
away from the Sun.
For this reason, we make the simple assumption that in 
a given GGC the MACHOs are all at the 
same distance from the Sun ($r_{sun}$, as given in Table \ref{GGC1}).
Table \ref{GGC2} shows, for the whole subset of inner events, the predicted deflector mass in units of solar masses, $<m>$, 
obtained with these assumptions.
The resulting average lens mass gets values in the range $\{10^{-2},10\}$, suggesting that the involved deflectors are possibly 
either brown dwarfs, M-stars or stellar remnants. Moreover, Jupiter-like deflectors are not definitively excluded, since, 
already a small increase on $D_{os}$ can substantially reduce the predicted lens mass.

The average expected lens mass has been drawn from the set of inner events,
some of which being possibly not due to GGC MACHOs.
This source of contamination should be removed before
one makes any prediction, but since we are not
able to do such a distinction, the average values on the overall inner sample can be
taken as a first crude approximation.

Also given in Table \ref{GGC2} is the number of expected events toward
the GGC centers, $\Gamma_{exp}$, as calculated through formula (36) of ~\cite{straessle}~,
where it is assumed that all the lenses have the same mass, $\mu_o$,
in units of solar masses and that their distribution is
very narrow with respect to that of the source population. $\Gamma_{exp}$ is given in units 
of $f\times \mu_o^{-1/2}\times 10^{-3}/year$,
$f$ being the fraction of dark matter (in form of brown dwarfs, dim stars or stellar remnants) in the cluster.
For a typical value of $10^2-10^3$ monitored source stars behind a GGC (this number depending also on the GGC extension)
and an observation period of $\sim$ 5 to 10 years, we expect at most between half an event and a couple
of events toward each GGC depending also on the value of $f$, in reasonable
agreement with the results of Table \ref{GGC2}.


Clearly, the expected
number of events, and thus the rate, is certainly quite small
so that more observations are needed.
A possible strategy would be to survey systematically during many years
the line of sight comprising the four globular clusters
which we analysed.
In spite of all the mentioned limitations,
we believe that our results, although not conclusive, suggest
that some events might indeed be due to lenses 
located in globular clusters.

\end{document}